# New design approach of Front-End Electronics for high-accuracy time measurement systems used in particle detection

**A. GHIMOUZ**[a,*]**, F. RARBI**[b] **and O. ROSSETTO**[b]

[a] *Univ. Grenoble Alpes, Grenoble INP, CNRS, LPSC-IN2P3,
38000 Grenoble, France*

[b] *Univ. Grenoble Alpes, CNRS, Grenoble INP, LPSC-IN2P3,
38000 Grenoble, France*
*E-mail* : ghimouz@lpsc.in2p3.fr

ABSTRACT: This paper discusses a detailed design approach to determine the optimal input impedance ($R_{in}$) and bandwidth (BW) for current preamplifiers in Front End Electronics (FEE) of high-accuracy time measurement systems used in particle detection. Our study shows the effect of the input impedance $R_{in}$ including the parasitic interconnection inductances of bonding wires between the detector and the FEE. We explain also the development of a new mathematical model for the estimation of the timing jitter of the current preamplifier in the case of using a low capacitor detector ($C_d$) as diamond. Different simulations based on developed MATLAB Simulink behavioral models are done in addition to an electric simulation of a transimpedance amplifier (TIA) designed in a 130 nm 1P8M CMOS technology which demonstrates the accuracy of the design approach and the timing jitter estimation model.

---

[*] Corresponding author.

# Contents

## 1. Introduction

Achieving high resolution time measurements using chemical vapor deposition (CVD) diamond represents an actual challenge for microelectronic design applied for particle detectors. Due to its outstanding performance: fast timing response: high charge mobility (2200 cm$^2$/Vs and 1600 cm$^2$/Vs for electron and hole respectively), low leakage current: high bandgap (5.45 eV), and so on as summarized in table I.

**Table I.** PROSPERITIES OF DIAMOND VS SILICON [1] - [2]

| Property | Units | Diamond | Silicon |
|---|---|---|---|
| Band Gap $E_g$ | eV | 5.47 | 1.12 |
| Electron mobility $\mu_e$ | cm$^2$ / V·s | 1700 | 1420 |
| Hole mobility $\mu_h$ | cm$^2$ / V·s | 2100 | 470 |
| Saturation velocity | cm / s | $2\times10^7$ | $1.4\times10^7$ |
| Intrinsic carrier density | cm$^{-3}$ | $<10^3$ | $1.5\times10^{10}$ |
| e/h pair energy | eV | 13 | 3.6 |
| Displacement energy | eV | 37-47 | 15-20 |
| Density | g cm$^{-3}$ | 3.52 | 2.33 |
| Rad length $X_0$ | cm | 12.2 | 9.4 |
| Dielectric constant $\varepsilon_r$ | (relative) | 5.7 | 11.9 |
| Breakdown E-Field | V/μm | 1000 | 30 |
| Resistivity | Ω/cm | $>10^{15}$ | $10^5$ - $10^6$ |

CVD diamond offers an attractive alternative to Si-PIN detector in nuclear physics experiment and pulse radiation research [3]. This creates a need for dedicated Front-end Electronics (FEE) design that guarantees a minimum impact on the generated signals: optimum bandwidth, low noise and linearity.

We define a time measurement system as a FEE for radiation detectors that targets a time to digital converter (TDC), figure 1 illustrates its different components.

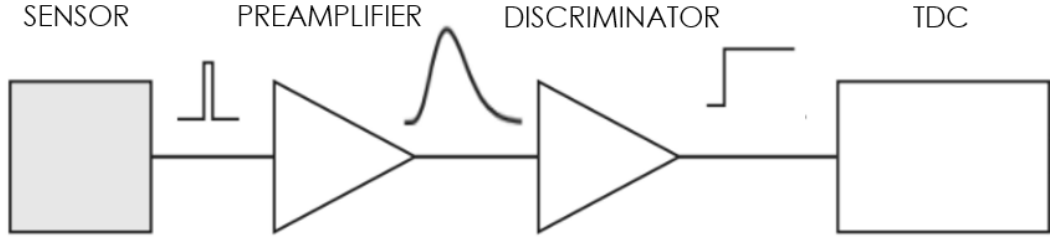

**Figure 1.** FEE diagram for radiation detectors

We focus on the input stage: the preamplifier, where we study the effect of its features on the final timing resolution. This latter is mathematically modeled as the maximum error of time defined using equation (1):

$$\sigma_t = \sqrt{\sigma_{Jitter}^2 + \sigma_{TimeWalk}^2 + \sigma_{TDC}^2} \qquad (1)$$

Where:
- $\sigma_{TDC}$ is the resolution of the TDC which depends on its architecture.
- $\sigma_{TimeWalk}$ is related to the discrimination stage, it appears when signals with same rise time but different amplitude are detected. There are many techniques to eliminate this error [4].
- $\sigma_{Jitter}$ is usually associated to the noise of the preamplifier stage which causes an uncertainty in the threshold crossing time.

The $\sigma_{Jitter}$ represents the most critical element which we need to reduce in order to achieve high time resolutions (<100 ps). During this study, this error will be considered as the criteria of time resolution.

Many works have been done in the literature [5-6] aiming to find the optimal possible values of the input impedance of the FEE so to get the best performances for high-resolution time measurement. Classical studies suggest that having the lowest possible value of the input impedance will limit the time constant associated to the input node of the FEE [5-6]. These design criteria propose also to maximize the FEE bandwidth BW, which is usually deemed as the dominant parameter affecting the dynamic performance of the read-out systems. A recent study [5-6] shows that, these criteria (i.e. low Rin and high BW) do not represent the best solution in real situations; where the characteristics of the detector and the parasitic coupling inductance are taken into consideration. It exists an optimum range of values for the input impedance and the FEE bandwidth [5-6].

In this paper, we developed a new mathematical model capable of obtaining the optimum values of Rin and BW of the FEE used for high-resolution time measurement (<100 ps). This allows us to make better design choices before going through many different electric simulations that usually take a lot of time and effort. The study is based on a design of a FEE for CVD diamond, which can detect a minimum charge of 5 fC from a pulse of few ns. The model is designed in MATLAB Simulink and compared to electrical simulations of our developed amplifier stage.

This paper is organized as follows: Section II introduces the system level modeling of the proposed CVD detector, the electronic stages and the analyses of several constraints in designing the FEE. Section III describes details of the implementation of the preamplifier. The simulation results of the preamplifier and the comparisons with the theoretical model are presented in Section IV followed by the conclusion in Section V.

## 2. System level modeling

### 2.1 CVD diamond detector characterization

The CVD diamond detector of this study is a double-side stripped metallized diamond used as a position sensitive detector. It is illustrated in figure 2.

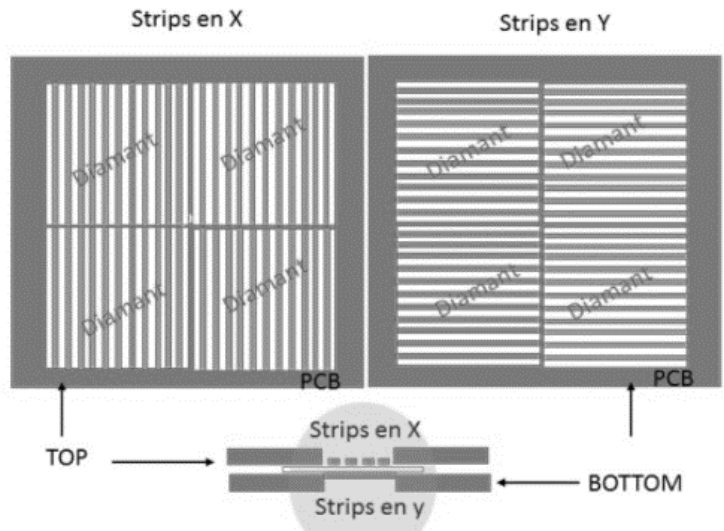

**Figure 2.** CVD striped diamond detector

In this case, diamonds are used as solid-state ionization chambers: the motion of the charges generated by the particles hitting the diamond creates an instantaneous current that induces on the electrodes connected to the strips. This means that the detector can be modeled as a current source with both a capacitor and a resistor in parallel of each other. We focus on the estimation of the equivalent capacity of the detector (i.e. for a couple of two strips ($x_i$, $y_i$)). The resistor is neglected because it is so high to be considered (several TΩ) [7]. Using the dimensions of the detector shown in figure 3, the simplest estimation of the equivalent capacity is:

$$C = \frac{A}{d} \times \varepsilon_0 \times \varepsilon_r \qquad (2)$$

Where:
$A$ is the surface of the cross-section of the strips
$d$ is the thickness of the detector
$\varepsilon_r$ is the electric permittivity of free space
$\varepsilon_0$ is the relative permittivity of the diamond

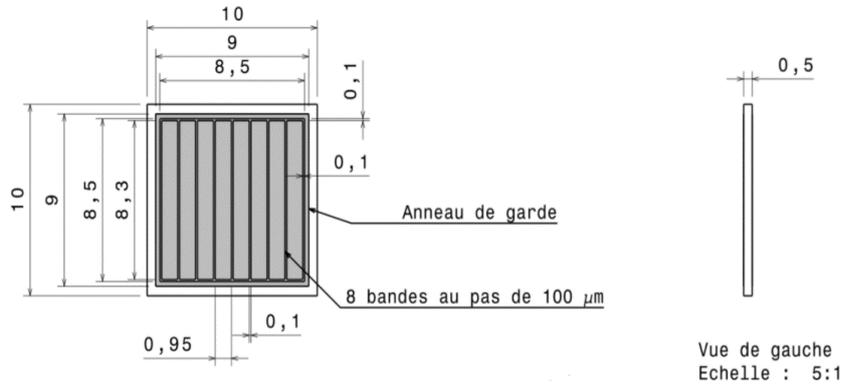

**Figure 3.** Diamond detector dimension

## 2.2 Electrical model of the system

The system used in this study is composed of two main parts: the electric equivalent model of the detector detailed above and the FEE represented by its input impedance. We assume in our model that the input impedance is represented by its resistive component. This will be explained in subsection 3.1 where we detail the expression of the developed amplifier.

We explore the effect of the value of the input impedance on the performance of the generated signal using a first model shown in figure 4. This can be seen as a first order system characterized by a time parameter $\tau$ as described in equation (3):

$$\tau = R \times C_d \quad (3)$$

A MATLAB Simulink model is developed to represent the response of the equivalent electric circuit of this model.

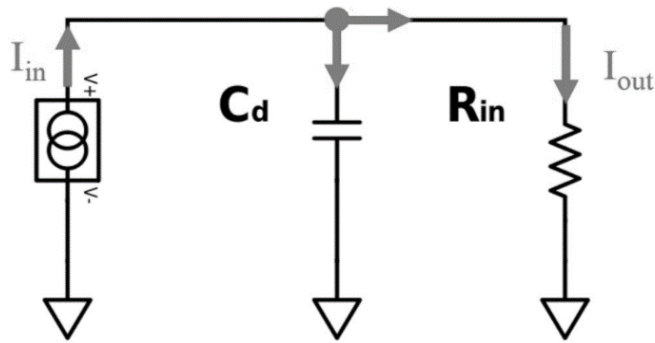

**Figure 4.** Electrical model of the input circuit

Using this model, we can confirm the classical criteria that suggest the smallest input impedance to reach the fastest response. Unfortunately, this is not true for a real system; when the input impedance reaches a small value ($R_{in}$ < 20 Ω), the system's response starts showing oscillations. This result indicates that the model used is not completely representing the real response of the system. A more accurate model is proposed adding the interconnection inductances of bonding wires between the detector and the FEE. It is represented in figure 5.

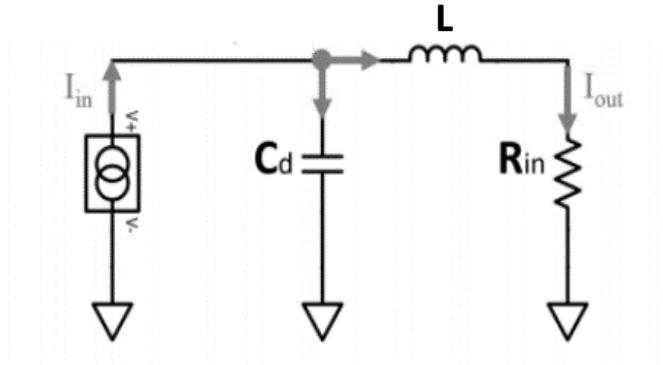

**Figure 5.** Precise electrical model of the input circuit with bonding wire inductance

Mathematically, this is modeled by the following transfer function:

$$H(p)_{\frac{I_{out}}{I_{in}}} = \frac{1}{1+(R_{in}\times C_d)\times p+(L\times C_d)\times p^2} \tag{4}$$

The response of a second order system depends on the value of the damping factor. In our case, we calculate this based on the classical model of a second order model:

$$H(p) = \frac{G}{1+\frac{2\times \xi}{\omega_0}\times p+\frac{1}{\omega_0^2}\times p^2} \tag{5}$$

Comparing equations (4) and (5), we find that the damping factor value is related to the value of the input impedance, the equivalent capacity of the detector and the value of the parasitic inductance of the bonding wires as shown in equation (6):

$$\xi = \frac{R_{in}\times C_d}{2\times\sqrt{L\times C_d}} \tag{6}$$

When a second order system oscillates, we call it underdamped ($\xi < 1$) which is explained as the existence of a pseudo angular frequency, its value is related to the damping factor as demonstrated in the following equation:

$$\omega_d = \omega_n \times \sqrt{1-\xi^2} \tag{7}$$

To ensure the stability of the system and to avoid oscillations, the value of the damping factor must be slightly greater than "1" which means that the input impedance has to be limited. This explains why the classical criteria is not working for real situations. The optimum value of the input impedance is the one that allows us to have a damping factor $\xi > 1$: a fast system without oscillations. To investigate this optimum value, we plotted in figure 6 the evolution of the damping factor for different values of input impedance and parasitic bonding wire inductance. The capacitance of the detector has the same value estimated in the previous section.

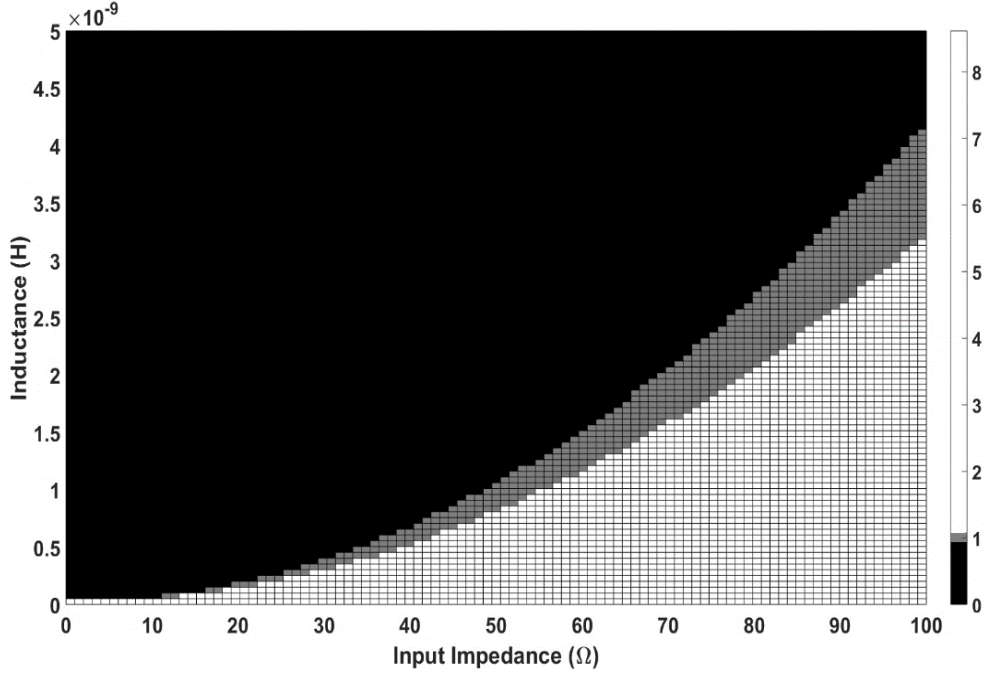

**Figure 6.** Values of the damping factor for different values of input impedance and parasitic inductance

The optimum choice of the input impedance is to take the value that gives the closest damping factor to 1 according to the value of the estimated parasitic inductance. This allows us to obtain the fastest response avoiding any oscillations.

**2.3 Bandwidth optimization**

One of the critical challenges in designing FEE for fast systems with high accuracy of timing measurement is to find the suitable bandwidth of the FEE. Usually, wideband systems are the most recommended [5-6]. However, this is the most difficult design when it comes to low noise and low power consumption efficiency. So, one can ask: is it truly necessary for fast systems with high accuracy of timing measurement to be wideband? In this section, an answer to this question is proposed based on the study of the RMS value of the timing jitter of the output FEE stage as a criterion of time performance.

When time is measured using discriminator systems, the timing jitter is classically defined as the ratio of the RMS output noise "$\sigma_n$" and the slope of the output signal "$\frac{dV}{dt}$" [8]:

$$\sigma_{Jitter} = \frac{\sigma_n}{\frac{dV}{dt}} \quad (8)$$

For fast systems, this value is also estimated as the ratio of the rise time of the output signal and the SNR of the system [8]:

$$\sigma_{Jitter} = \frac{T_{rise}}{SNR} \quad (9)$$

This equation shows that in order to minimize the timing jitter error, we need to minimize the rise time of the system to match the rise time of the detector and maximize the SNR.

For a preamplifier, the rise time is related to its bandwidth as illustrated in the following equation:

$$T_{rise} = \frac{0.35}{BW} \tag{10}$$

From this, we notice why it is classically recommended to choose wideband systems. The main issue in this choice is the fact that increasing bandwidth will lead to an increasing of the noise level, which will therefore decrease the SNR. Thus, the timing jitter will be degraded. We propose here to study a new kind of expression of the timing jitter equation with reference to equation (8).

The first step is to define the expression of the RMS output noise. In the case of a TIA preamplifier. The equivalent output noise can be estimated using three parameters: the $BW$, the current input noise $i_n$ and the gain of the amplifier $G$ as shown in equation (11):

$$\sigma_n = G \times i_n \times \sqrt{\alpha \times BW} \tag{11}$$

Where:
  $\alpha$ is the multiplication factor of the Equivalent Noise Bandwidth (ENBW) [9]

The second step consists of approaching the value of the slope of the output signal as the ratio of the maximum value of the output signal and its rise time, this approximation is the same used for equation (9):

$$\frac{dV_0}{dt} = \frac{V_{max}}{T_{rise}} \tag{12}$$

The maximum output of a TIA amplifier with a reference to its output is:

$$V_{max} = G \times I_{in_{max}} \tag{13}$$

The rise time at the output of the amplifier is defined by:

$$T_{rise} = \sqrt{T_{r1}^2 + T_{r2}^2 + T_{r3}^2} \tag{14}$$

Where:
  - $T_{r1}$ is the rise time of the detector signal.
  - $T_{r2}$ is the rise time of the input stage.
  - $T_{r3}$ is the rise time of the preamplifier stage.

Using the values of equation (11) and (12) in equation (9), we find a new mathematical equation for the timing jitter:

$$\sigma_{Jitter} = \frac{i_n \times \sqrt{\alpha \times BW} \times T_r}{I_{in_{max}}} \tag{15}$$

This new equation shows that the electronics is not the only factor that influence the value of the jitter. The performances of the detector do too (The current generated by the detector $I_{in_{max}}$ that depends on its characteristics) but since this later is so difficult to adjust, we focus only on the design choices of the electronics.

First, we studied the effect of having a wide $BW$ system on the value of its timing jitter as suggested in the classical criteria. The plot of the evolution of the timing jitter for different values of $BW$ is illustrated in figure 7. The timing jitter function is calculated for a system with the following values:

- $i_n = 13.2\ pA/\sqrt{Hz}$ (This means that we manage to keep the maximum of the current input noise $i_n$ below or equal to this value for every chosen $BW$).
- $T_{r1} = 350\ ps$.
- $T_{r2} = 100\ ps$.
- $T_{r3} = \frac{0.35}{BW}$;
- $I_{in_{max}} = 5\ \mu A$.

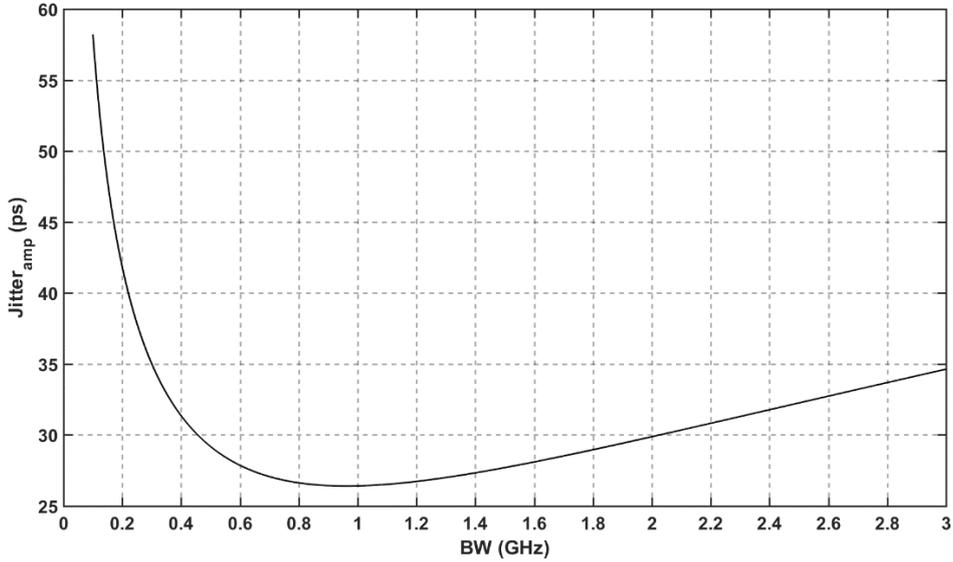

**Figure 7.** Timing jitter function according to FEE

Analyzing results of figure 7, we can clearly conclude that the effect of the BW on the timing jitter is considerably weak. If we compare a system with a BW of 2 GHz with a system of 500 MHz, we have barely an improvement of around 10 ps for a higher price of complexity in design (maintaining a low value of the current input noise $i_n$). This figure shows also the existence of an optimum value of BW for the lowest possible timing jitter. This optimum represents the minima (value of the BW) of the function of timing jitter equation (15). We identify the value of it using the second derivative test and we found:

$$BW_{optimum\ Jitter} = \frac{0.35}{\sqrt{T_{r1}^2 + T_{r2}^2}} \qquad (16)$$

These results leaded us to our second study, where we analyzed the evolution of the value of $T_{rise}$ of the system which depends on the choice of the BW as shown in equation (2): Having wide BW system will make the value of $T_{r3}$ too close to the value of $T_{r1}$ which is related to the input stage. This means that this stage will be more critical in the calculation of the final value of $T_{rise}$. As shown in the modeling section (subsection 2.2), the input stage is represented as a second order system, since we guarantee the overdamped case, the rise time of the input stage $T_{r2}$ is estimated as shown in equation (17)[10]:

$$T_{r2} = \frac{1.8}{\omega} \tag{17}$$

Where:

$$\omega = \frac{2 \times \pi}{\sqrt{L \times C_d}} \tag{18}$$

Thus, in the case of having a wide BW, the value of the parasitic inductance must be taken into consideration in the estimation of the final resolution since there is a strong correlation.

From these two analyses, we can clearly identify an optimum BW for the targeted value of the timing jitter while minimizing the impact of the input stage.

## 3. Circuit implementation

In order to validate the new timing jitter expression: equation (15), a TIA was designed [11]. We show in this section the TIA architecture and its several specifications.

### 3.1 TIA architecture

The proposed TIA architecture is based on a common gate feedforward structure [11] as illustrated in figure 8.

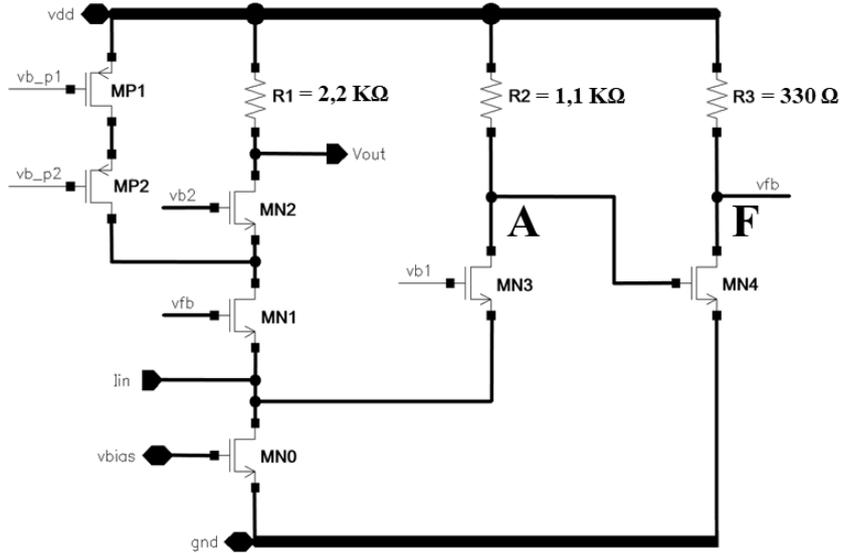

**Figure 8.** Common Gate Feedforward TIA

It is a common gate topology using a gain-enhancing feedforward path composed by the transistors MN2 and MN3. The transistor MN1 provides the feedback path for the current amplification so that the transistors MP1 and MP2 will permit to boost the gain amplification by reducing the current flow through the resistor R1. The transistor MN0 is used as the current source to biasing the full amplifier.

In low power supply design, it is difficult to bias all transistors in saturated region. Therefore, in our design, the transistor MN3 is inserted to shift the gate voltage of the transistor MN4 to a higher level. Even at 1.2V power supply, all amplifying transistors are biased at a gate-to-source and drain-to-source voltage at least equal to 0.5V.

The transimpedance low frequency gain is provided by the resistance R1 as:

$$\frac{V_{out}}{I_{in}} \approx R_1 \tag{19}$$

The input impedance of the TIA ($Z_{in}$) varies depending on the frequency, it can be described as following:

$$Z_{in} = \frac{V_{in}}{I_{in}} = \frac{1}{g_{m1} \times |A_1(f)| \times |A_2(f)|} \tag{20}$$

With $A_1(f)$ and $A_2(f)$ are respectively the gains the second and third stage of the amplifier. The second stage is a common gate amplifier and the third stage is a common source amplifier:

$$A_1(f) = \frac{g_{m3}.R_2}{1+R_2.C_A.p} \tag{21}$$

$$A_2(f) = \frac{-g_{m4}.R_3}{1+R_3.C_F.p} \tag{22}$$

Where:
$C_A$ and $C_F$ are the equivalent capacitors in the node A and F shown in figure 8
$g_{mi}$ is the trans-conductance of transistor MN$_i$

Replacing equation (21) and (22) in equation (20) give us:

$$Z_{in} = \frac{(1+R_2.C_A.p) \times (1+R_3.C_F.p)}{g_{m1} \times (R_2 \cdot g_{m3} \cdot R_3 \cdot g_{m4})} \tag{23}$$

This expression (equation 23) shows that $Z_{in}$ has an imaginary component (which is proportional to the trans-impedance gain and the bandwidth of the amplifier) represented by the two zeroes $R_2.C_A$ and $R_3.C_F$ which we can notice clearly on figure 9 (black dashed line). Yet, these two zeros take effect in high frequencies. We work in lower frequencies where the input impedance is only represented by its resistive component:

$$Z_{in} \approx \frac{1}{g_{m1} \times (R_2 \cdot g_{m3} \cdot R_3 \cdot g_{m4})} \tag{24}$$

This explained clearly that we represented the input impedance as a resistor ($R_{in}$) in our theoretical model (subsection 2.2).
Using this kind of topologies, allows us to get very low input impedance via the use of $R_2$ and $R_3$ (i.e., few ohm).

**3.2 TIA features**

The bode diagram (black line), the input referred noise spectrum (grey point-dashed line) and input impedance (black dashed line) are illustrated in figure 9. The low frequency gain is about 66 dBΩ and the cutoff frequency is approximately equal to 560 MHz. The input referred noise at the noise frequency: $\frac{\pi}{2} \times f_{-3dB}$ [9] is estimated about 13.2 pA/$\sqrt{Hz}$ and we have an RMS integrated input referred noise of 450 nA$_{RMS}$. All of these parameters will be used to estimate the expected timing jitter of our amplifier.

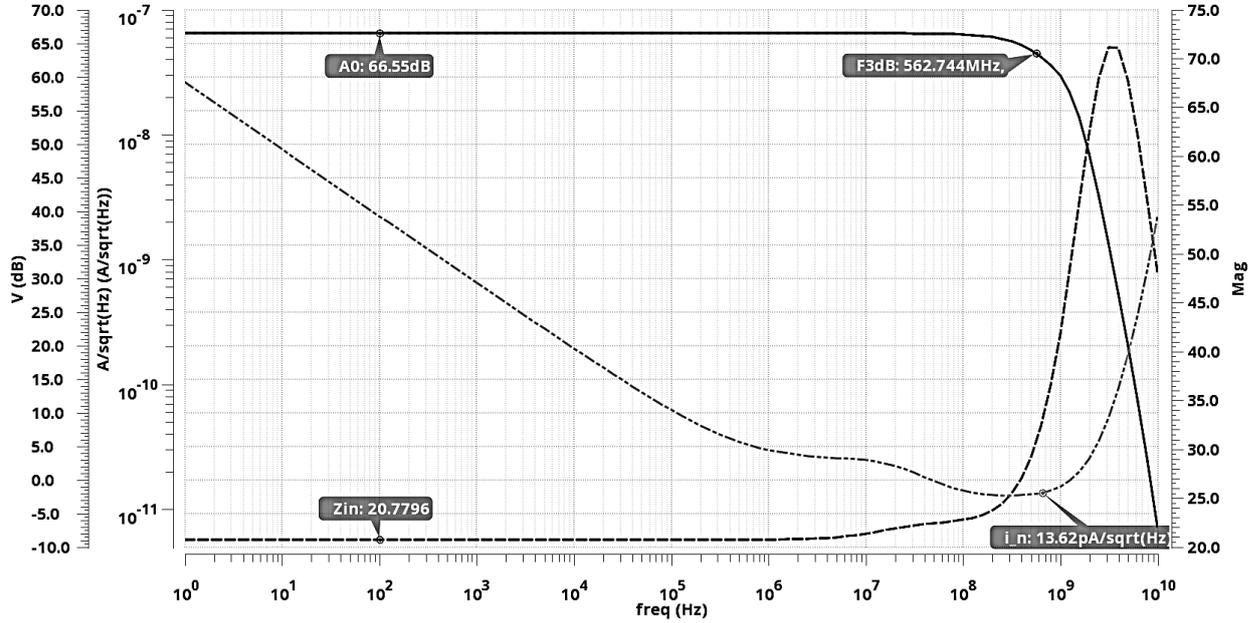

**Figure 9.** AC and noise extracted simulation results

The key performance parameters of the common gate feedforward TIA are summarized in table II.

**Table II.** Key performance parameters

| Parameters | Extracted simulation results |
|---|---|
| TIA Gain (dBΩ) | 66.55 (2125 V/A) |
| DC input impedance (Ω) | 20.78 |
| TIA Bandwidth ($f_{-3dB}$ – MHz) | 562.74 |
| Input-ref. Noise (pA/$\sqrt{Hz}$) @ $\frac{\pi}{2} \times f_{-3dB}$ | 13.62 |
| Power consumption @1.2V (mW) | 2.54 |

## 4. Simulation results

In this section, we illustrate by extracted simulation results the several issues of a very low input impedance (Zin). As mentioned earlier in section II-B, bonding wires can deteriorate the response of the TIA in accordance with low impedance. Figure 10 shows oscillation issues with a Rin of 20Ω and several inductance values (5nH, 10nH and 20nH).

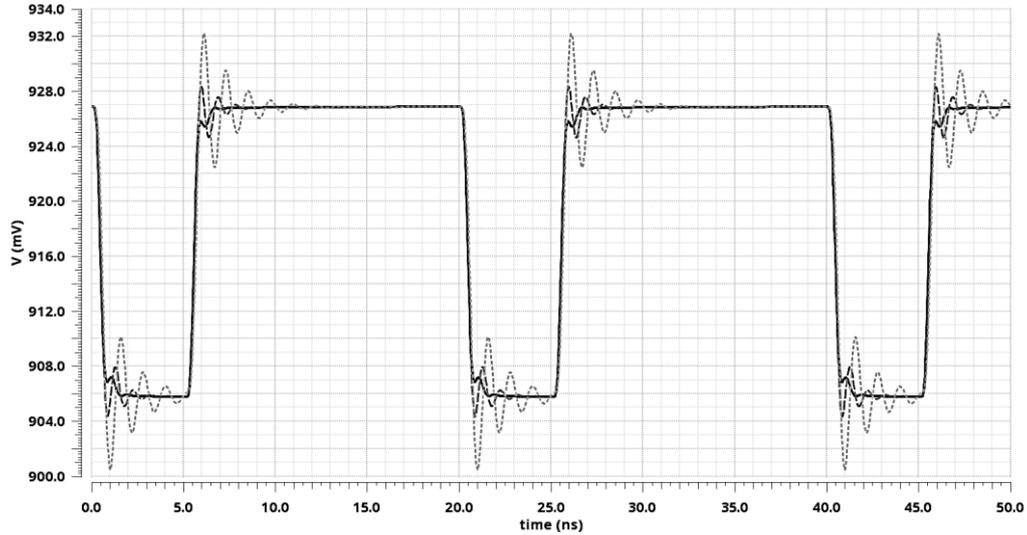

**Figure 10.** Extracted Transient simulation of the TIA output with 20Ω input impedance and several inductance values (5nH (solid), 10nH (dashed) and 20nH (doted))

To reduce this oscillation phenomena, we have to adapt the signal path and/or increase the input impedance. Unfortunately, in this last case the amplifier bandwidth will decrease. Figure 11 shows the case with an input impedance Rin of 50Ω: the bandwidth of the TIA is reduced to approximately 400MHz.

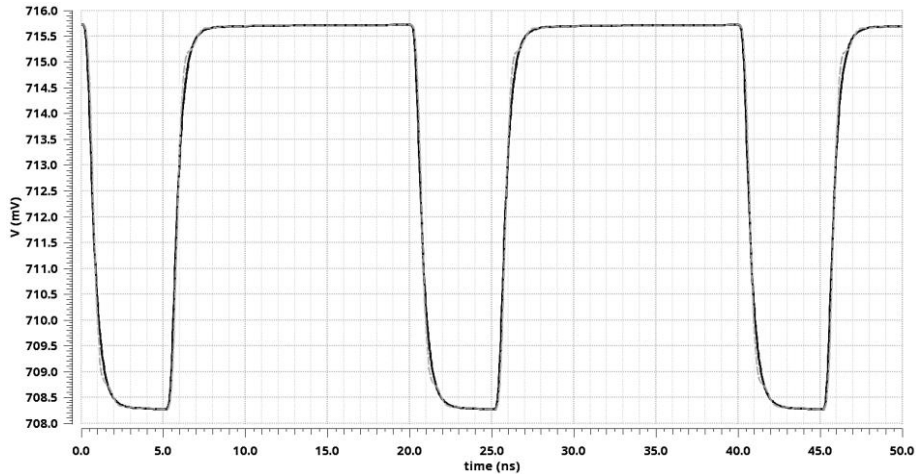

**Figure 11.** Extracted Transient simulation of the TIA output with 50Ω input impedance and several inductance values (5nH, 10nH and 20nH)

In a second time, we compared the extracted simulation results of the timing jitter with both the classical formula (equation (9)) and our new formula (equation (15)). The simulated results are obtained from RC-extracted transient noise simulation and the estimation of the eye diagram of the output signal. These results are illustrated in figure 12. We achieved around 50 ps of timing jitter using the designed preamplifier with very low input current. The new formula shows more accurate results than the classical one compared to the extracted simulations. In order to show that more precisely, we calculated the error between both formulas and the simulations.

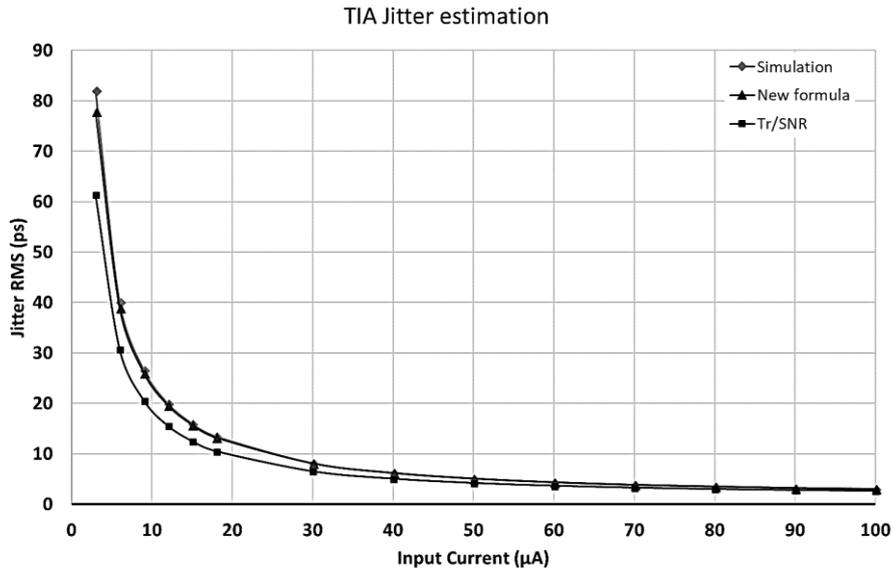

**Figure 12.** The values of the timing jitter: simulated and calculated

Figure 13 shows that the new formula is more accurate than the classical one: Using the classical formula, we have an error that varies between 15% and 25 % while the new formula has less than 5% error.

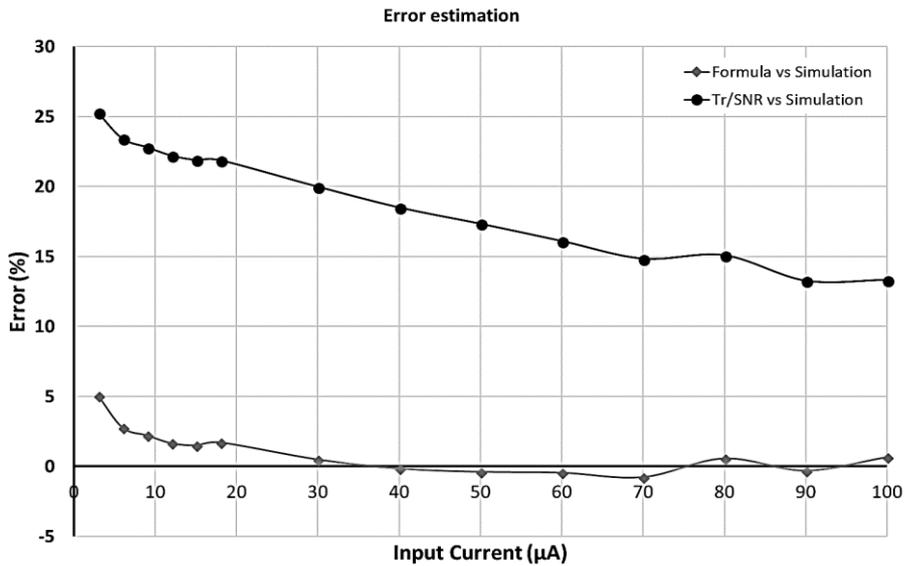

**Figure 13.** The values of the timing jitter: simulated and calculated

## 5. Conclusion

In this paper, we discussed and analyzed a design approach of FEE electronics for low capacity detectors. More precisely, a method to define the optimal input impedance and bandwidth for current preamplifiers in FEE of high-accuracy time measurement systems used in particle detection. The following results are reported here:

- We developed a detailed approach to define precisely the optimal input impedance taking in consideration the impact of the parasitic interconnection inductances of bonding wires between the detector and the FEE in order to have the fastest response avoiding any kind of oscillations.
- We explained the development of a new mathematical model for the estimation of the timing jitter of the current preamplifier in the case of using a low capacitor detector (like diamond) and we used it to calculate the optimal bandwidth of the preamplifier. This study showed also the effect of the input stage and the detector parameters on the timing accuracy.
- We tested in the first time the developed mathematical models using MATLAB Simulink models then we implemented it using a TIA designed in a 130 nm 1P8M CMOS technology.
- We demonstrated the accuracy of our models through electric simulation results of the TIA.

These results prove the accuracy the detailed approach and gives more details about the calculation of the optimal input impedance and the bandwidth of current preamplifiers in FEE used in high-accuracy time measurement systems. Further details will be discussed after testing the ASIC that is currently under fabrication.

**Acknowledgments**

We thank BB130 – IN2P3 Collaboration for financial and technical supports.